\documentclass[10pt,journal,epsfig]{IEEEtran}
\usepackage{graphicx}
\usepackage{subfigure,color}
\usepackage{amssymb}
\usepackage{cite,comment,setspace,flushend}
\usepackage{amsmath}
\usepackage{bm,comment}
\usepackage{algorithm}
\usepackage{algpseudocode}
\makeatletter

\newcommand{\Rmnum}[1]{\expandafter\@slowromancap\romannumeral #1@}
\newcommand{\mv}[1]{\mbox{\boldmath{$ #1 $}}}

\makeatother

\newtheorem{remark}{\underline{Remark}}

\begin{document}
\title{Distributed Beam Training for Intelligent Reflecting Surface Enabled Multi-Hop Routing}
\author{Weidong Mei and Rui Zhang, \IEEEmembership{Fellow, IEEE}
\thanks{\footnotesize{The authors are with the Department of Electrical and Computer Engineering, National University of Singapore, Singapore 117583 (e-mails: \{wmei, elezhang\}@nus.edu.sg).}}}
\maketitle

\begin{abstract}
Intelligent reflecting surface (IRS) is an emerging technology to enhance the spectral and energy efficiency of wireless communications cost-effectively. This letter considers a new multi-IRS aided wireless network where a cascaded line-of-sight (LoS) link is established between the base station (BS) and a remote user by leveraging the multi-hop signal reflection of selected IRSs. As compared to the conventional single-/double-hop IRS system, multi-hop IRS system provides more pronounced path diversity and cooperative passive beamforming gains, especially in the environment with dense obstacles. However, a more challenging joint active/passive beamforming and multi-hop beam routing problem also arises for maximizing the end-to-end channel gain. Furthermore, the number of IRS-associated channel coefficients increases drastically with the number of IRS hops. To tackle the above issues, in this letter we propose a new and efficient beam training based solution by considering the use of practical codebook-based BS/IRS active/passive beamforming without the need of explicit channel estimation. Instead of exhaustively or sequentially searching over all combinations of active and passive beam patterns for each beam route, a distributed beam training scheme is proposed to reduce the complexity, by exploiting the (nearly) time-invariant BS-IRS and inter-IRS channels and the cooperative training among the BS and IRSs' controllers. Simulation results show that our proposed design achieves the end-to-end channel gain close to that of the sequential beam search, but at a much lower training overhead and complexity.
\end{abstract}

\section{Introduction}
Achieving smart and controllable wireless environment is a desired goal of the next generation wireless communication network, which can be realized by the emerging intelligent reflection surface (IRS) technology\cite{wu2020intelligent}. Specifically, IRS is a two-dimensional planar surface consisting of a large number of programmable reflecting elements, which can be jointly tuned to alter the strength and direction of an impinging electromagnetic wave, thereby reconfiguring the wireless channel to be more favorable for communication. Since the reflecting elements are passive, they drastically reduce the hardware cost and power consumption of traditional active arrays.

\begin{figure}[!t]
\centering
\includegraphics[width=3.2in]{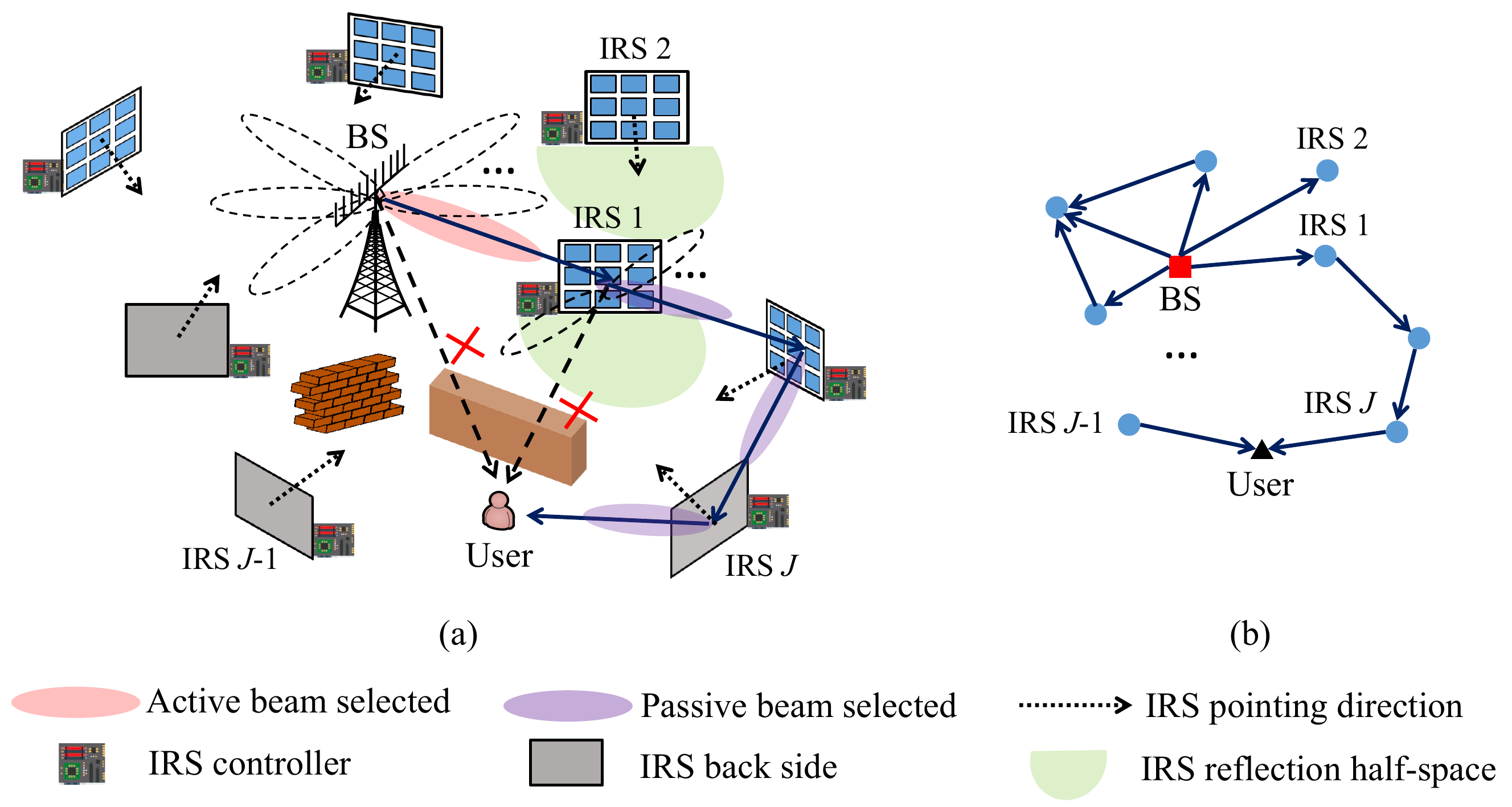}
\DeclareGraphicsExtensions.\vspace{-6pt}
\caption{(a) A multi-IRS aided wireless network with multi-hop beam routing; (b) the corresponding LoS link graph.}\label{MultiBeam}
\vspace{-15pt}
\end{figure}

The great potential of IRS has aroused an intensive interest in investigating its design and performance in various wireless systems and applications\cite{wu2020intelligent,basar2019wireless}. However, most of the existing works on IRS considered the IRS-aided wireless systems with only one single reflection by one or multiple distributed IRSs, while the potential inter-IRS reflections were ignored. Note that multi-hop IRS reflection links can provide more path diversity to bypass the dense and scattered obstacles in a complex environment (e.g., indoor), as shown in Fig.\,\ref{MultiBeam}(a). Moreover, they can offer more pronounced cooperative passive beamforming (CPB) gain than single-hop IRS reflection links, by exploiting the inter-IRS line-of-sight (LoS) links, which also effectively overcomes the ``product'' path-loss that generally increases with the number of IRS hops. Motivated by the above advantages, a handful of recent works have investigated the power scaling law\cite{han2020cooperative}, optimal CPB design\cite{zheng2021double,dong2021double}, and channel estimation methods\cite{you2021wireless,zheng2021efficient} for the double-IRS aided wireless system with both single-hop and double-hop signal reflections. 

However, several new challenges arise in the general multi-IRS aided wireless network with multi-hop (more than two hops) signal reflections, as shown in Fig.\,\ref{MultiBeam}(a). First, there usually exist multiple multi-IRS/hop reflection links between a BS and its served user, which gives rise to a new path selection and cooperative beam routing design problem, aiming to find the optimal multi-hop reflection path and design the active beamforming at the BS jointly with the passive beamforming at all selected IRSs in the optimal path to maximize the end-to-end channel power gain between the BS and the user. Furthermore, for each reflection path, the associated BS/IRS active/passive beamforming design requires channel state information (CSI) on all of its constituent BS-IRS, inter-IRS, and IRS-user links, which is difficult to obtain in practice, especially when the number of IRS hops becomes large. In fact, existing methods for estimating the BS-IRS-user cascaded channel\cite{wu2020intelligent}, as well as those extended to the double-IRS system\cite{you2021wireless,zheng2021efficient} are generally not applicable to the multi-IRS system with more than two hops. On the other hand, there have been some initial studies\cite{mei2020cooperative,mei2021massive,huang2021multi} on the BS/IRS active/passive beamforming design for the multi-hop IRSs. However, they all assumed perfect CSI on all links. 

In this letter, we aim to tackle the above practical challenges for the multi-IRS system with multi-hop signal reflection, by exploiting a new approach based on distributed active/passive beam training at the BS/IRSs. We consider a practical wireless system where the BS and each IRS employ a predefined active/passive beamforming codebook, which consists of only a finite number of beam directions or patterns, as depicted in Fig.\,\ref{MultiBeam}(a). However, for a given multi-hop reflection path, its beam training requires an exhaustive or sequential search over all possible combinations of beam patterns at the BS and involved IRSs, which incurs excessively high training complexity and time overhead, especially when the hop number and/or the resolution of beamforming codebooks increases. To resolve this issue, we propose a new beam training method with combined offline and online distributed beam training, by exploiting the (nearly) time-invariant BS-IRS and inter-IRS channels and the cooperative training among the BS and IRSs' controllers. In particular, the BS and all IRSs create local beam routing tables (BRTs) and exchange them with each other to facilitate the optimal multi-hop path selection and active/passive beamforming design. Based on the global BRT at the BS, an efficient beam-routing solution is proposed, jointly with the active/passive beamforming design for the BS/selected IRSs. Simulation results show that our proposed distributed beam training based design yields the end-to-end channel gain comparable to that with the sequential beam search, but at a much lower training overhead and complexity.

\section{System Model}
As shown in Fig.\,\ref{MultiBeam}, we consider the downlink transmission in a multi-IRS aided wireless system, where a multi-antenna BS communicates with a single-antenna user with the help of $J$ densely distributed IRSs, denoted as ${\cal J}\triangleq \{1,2,\cdots,J\}$. For convenience, we refer to the BS and the user as nodes 0 and $J+1$, respectively. Due to dense obstacles in the complex environment (e.g., indoor), the BS can communicate with the user efficiently only via a multi-hop LoS link formed by a set of selected IRSs, as shown in Fig.\,\ref{MultiBeam}, which is much stronger than the other (randomly) scattered links between them. We assume that the BS is equipped with $N$ antennas, while each IRS is composed of $M=M_1 \times M_2$ reflecting elements, where $M_1$ and $M_2$ denote the numbers of elements in its horizontal and vertical dimensions, respectively. For convenience, we also assume that each IRS is placed perpendicular to the ground (see Fig.\,\ref{MultiBeam}). As each IRS can only achieve 180$^\circ$ half-space reflection, i.e., only the signal incident from its front half-space can be reflected, we define a pointing direction for each IRS which is perpendicular to its surface and points to its reflection half-space, as shown in Fig.\,\ref{MultiBeam}. 

Let $d_{i,j}, i \ne j$ denote the distance between nodes $i$ and $j$, for which some reference elements can be selected at the BS/IRSs for convenience. Moreover, a smart controller is attached to each IRS for tuning its passive reflection and exchanging control information with the BS and other IRS controllers via separate wireless links. We label the controller of IRS $j$ as IRS controller $j$. Each IRS controller $j$ is assumed to be located at the reference point of IRS $j$, such that its channels with the BS, other nearby IRSs as well as its served user (if any) can well approximate those of IRS $j$'s reference element with them. For example, in the case that the BS-IRS $j$ channel is dominated by an LoS path under the condition of far-field propagation, the channel from the BS to IRS controller $j$ is approximately equal to the array response of the BS in the direction towards IRS $j$. Nonetheless, this may not be achieved if there exists non-LoS (NLoS) propagation between them.

Let ${\mv w}_B \in {\mathbb C}^{N \times 1}$ and ${\mv \theta}_j = [e^{j\theta_{j,1}},e^{j\theta_{j,2}},\cdots,e^{j\theta_{j,M}}]^T \in {\mathbb C}^{M \times 1}, j \in \cal J$ denote the active beamforming vector at the BS and the passive beamforming vector at IRS $j$, respectively, where $\theta_{j,m}, m =1,2,\cdots,M,$ denotes the reflection phase shift of the $m$-th element of IRS $j$. Here we have set the reflection amplitude of each IRS element to its maximum value (i.e., unity). Thus, the reflection matrix of IRS $j$ is given by ${\mv \Phi}_j={\rm diag}({\mv \theta}_j), j \in \cal J$. Furthermore, for ease of implementation, we assume that the BS and each IRS select their active and passive beamforming vectors from their respective predefined codebooks, denoted as ${\cal W}_B$ and ${\cal W}_I$, i.e., ${\mv w}_B \in {\cal W}_B$ and ${\mv \theta}_j \in {\cal W}_I, \forall j \in \cal J$. Moreover, each IRS employs three-dimensional (3D) passive beamforming, where its passive beamforming vector can be decomposed as ${\mv \theta}_j={\mv \theta}^{(1)}_j \otimes {\mv \theta}^{(2)}_j, j \in \cal J$, where ${\mv \theta}^{(1)}_j \in {\mathbb C}^{M_1 \times 1}$ and ${\mv \theta}^{(2)}_j \in {\mathbb C}^{M_2 \times 1}$ denote its horizontal and vertical passive beamforming vectors, respectively, and $\otimes$ denotes the Kronecker product. Accordingly, we define ${\cal W}^{(1)}_I$ and ${\cal W}^{(2)}_I$ as the codebooks for the horizontal and vertical IRS passive beamforming, respectively, with ${\mv \theta}^{(1)}_j \in {\cal W}^{(1)}_I$ and ${\mv \theta}^{(2)}_j \in {\cal W}^{(2)}_I, \forall j \in \cal J$. Let $D_B$, $D_I$, $D^{(1)}_I$ and $D^{(2)}_I$ denote the numbers of beam patterns in ${\cal W}_B$, ${\cal W}_I$, ${\cal W}^{(1)}_I$ and ${\cal W}^{(2)}_I$, respectively, with $D_I=D^{(1)}_ID^{(2)}_I$.

For any two nodes $i$ and $j$, if they are both IRSs, each of them needs to be located in the reflection half-space (specified by the pointing direction) of the other to achieve effective signal reflection between them. For example, in Fig.\,\ref{MultiBeam}, IRS 1 and IRS 2 cannot successively reflect the signal from the BS as they do not meet the above condition. Similarly, if one of the two nodes (say, node $i$) is the BS/user and the other node (say, node $j$) is an IRS, then node $i$ should be located in the reflection half-space of node $j$ for achieving effective signal reflection by the IRS. Moreover, we consider that the BS's signal can be effectively reflected from one IRS to a farther IRS from the BS, but not vice versa, to minimize the propagation delay and path loss. 

Since the locations of the BS and all IRSs are fixed, the BS-IRS and inter-IRS channels can be assumed to remain constant over a long period\footnote{Although these channels may vary due to mobile scatterers, the LoS component in them (if any) is usually dominant in practice.}. As such, it is assumed that the LoS availability between any two nodes (IRS/BS) has been known {\it a priori} via the coordination among the BS and different IRS controllers based on existing methods (e.g., \cite{xiao2014non}). However, the LoS availability between an IRS and the user should be determined in real time (online) due to the user mobility. Accordingly, we construct a directed {\it LoS graph} for all the nodes and their wireless links in the system, denoted as $G_L=(V_L,E_L)$, where $V_L=\{0,1,\cdots,J\!+\!1\}$ denotes the set of vertices in $G_L$ and $E_L$ denotes the set of edges in $G_L$. In particular, we add an edge from vertex $i$ to vertex $j$, denoted as $e_{i,j}$, if the following two conditions are satisfied: 1) there is an LoS path between nodes $i$ and $j$ and effective signal reflection can be achieved between them, and 2) $d_{0,j} > d_{0,i}$ (for routing signal outwards from the BS) except that node $j$ is the user. For example, the LoS graph of the multi-IRS system in Fig.\,\ref{MultiBeam}(a) is shown in Fig.\,\ref{MultiBeam}(b). Thus, the set of neighboring nodes of IRS $j$ is denoted as ${\cal N}_j={\cal N}^{(P)}_j \cup {\cal N}^{(N)}_j$, where ${\cal N}^{(P)}_j=\{i|e_{i,j} \in E_L\}$ and ${\cal N}^{(N)}_j=\{r|e_{j,r} \in E_L\}$ are the sets of previous and next hopping nodes of IRS $j$ (or its predecessors and successors in the LoS graph), respectively. Similarly, we define ${\cal N}_0=\{j|e_{0,j} \in E_L\}$ as the set of next hopping nodes of the BS. Moreover, we define $Q_{i,j}$ as the average channel power gain between nodes $i$ (IRS controller) and node $j$ (IRS controller/user), with $e_{i,j} \in E_L$, which is measured offline/online for the case without/with the user involved. In practice, if an LoS link exists between nodes $i$ and $j$, $Q_{i,j}$ should be larger than a certain threshold\cite{xiao2014non}. We assume that all $Q_{i,j}$'s between any two IRS controllers $i$ and $j$ (with $e_{i,j} \in E_L$) have been measured offline and fed back to the BS by the corresponding controllers.

Define ${\mv H}_{0,j} \in {\mathbb C}^{M \times N_B}, j \in {\cal J}$ as the channel from the BS to IRS $j$, ${\mv g}_{j,J+1}^{H} \in {\mathbb C}^{1 \times M}, j \in {\cal J}$ as that from IRS $j$ to the user, and ${\mv S}_{i,j} \in {\mathbb C}^{M \times M}, i,j \in {\cal J}, i \ne j$ as that from IRS $i$ to IRS $j$. Note that each channel above depends on path-loss and small-scale fading, which may consist of both LoS and NLoS components. Given $J$ IRSs in total, different multi-hop LoS links are generally available from the BS to the user by selecting different subsets of intermediate IRSs from $\cal J$. Let $\Omega=\{a_1,a_2,\cdots,a_L\}$ denote a cascaded LoS reflection path from the BS to the user, where $L \ge 1$ and $a_l \in \cal J$ denote the number of intermediate IRSs in $\Omega$ and the index of the $l$-th intermediate IRS, with $l \in {\cal L} \triangleq \{0,1,2,\cdots,L\}$, respectively. For convenience, we define $a_0=0$ and $a_{L+1}=J+1$, corresponding to the BS and the user, respectively. Then, it follows that $e_{a_l,a_{l+1}} \in E_L, \forall l \in \cal L$ and the multi-hop BS-user channel under $\Omega$ is expressed as
\begingroup
\allowdisplaybreaks
\begin{equation}\label{recvsig}
h_{0,J+1}(\Omega)={\mv g}^H_{a_L,J+1}{\mv \Phi}_{a_L}\Big(\prod\limits_{l \in {\cal L}, l \ne L}{\mv A}_l\Big){\mv H}_{0,a_1}\mv w_B,\vspace{-3pt}
\end{equation}
where ${\mv A}_l={\mv S}_{a_l,a_{l+1}}{\mv \Phi}_{a_l}$, and we assume that all IRSs which are not included in $\Omega$ are switched off. From (\ref{recvsig}), the end-to-end channel gain depends on both the BS/IRS beamforming and the channels of its constituent links. As such, to attain the maximum gain of any BS-user multi-hop LoS channel, $h_{0,J+1}(\Omega)$, and thereby find the optimal beam routing solution or reflection path of $\Omega$ which achieves the maximum power gain among all multi-hop channels, all BS-IRS, inter-IRS, and IRS-user channels in the system need to be known, if they constitute an LoS link. However, this will incur prohibitively high complexity for channel training/estimation, if $J$ and/or $M$ is practically large. Therefore, we consider the practical approach that the BS and each IRS employ predefined active/passive beamforming codebooks, each with a finite number of fixed beam patterns. By proper beam training, the BS/IRS active/passive beamforming and beam routing can be optimized without the need of explicit channel estimation. 
\begin{remark}
It is worth noting that in addition to the selected multi-hop LoS link $\Omega$ from the BS to the user, there exist other signal paths in practice due to the uncontrolled (random) scattering among IRSs in $\Omega$ and by other environment objects. However, as will be shown in Section \ref{sim}, the strength of such randomly scattered links is practically much lower than that of the selected multi-hop LoS link $\Omega$ due to the lack of CPB gain over them, as the BS/IRS active/passive beamforming is designed to maximize the CPB gain in $\Omega$ only, which compensates for its multiplicative path-loss. As such, in this letter, we focus on maximizing the power gain of $h_{0,J+1}(\Omega)$ by ignoring other randomly scattered links in our design.\vspace{-6pt}
\end{remark}

\section{Exhaustive and Sequential Beam Search}\label{opt.train}
Based on (\ref{recvsig}), the optimal active/passive beamforming (${\mv w}_B$ and $\{{\mv \Phi}_{a_l}\}_{l=1}^L$) and beam routing ($\Omega$) that maximize the power gain of $h_{0,J+1}(\Omega)$ can be obtained based on the following beam training procedures.
In particular, for a given feasible path $\Omega$, to determine its corresponding optimal active/passive beamforming at the BS and all involved IRSs, the BS should coordinate with all IRS controllers in $\Omega$ to traverse all possible combinations of the active and passive beam patterns by training. In the meanwhile, the user measures its received signal strength (RSS) for each combination and reports the best one that achieves the strongest RSS to the BS. Then, similar procedures are conducted for other feasible reflection paths, until all feasible paths are searched. By this exhaustive search, the BS can determine the optimal beam routing solution (jointly with the associated best combination of beam patterns) by comparing the user's feedback RSS for all feasible paths. However, the total number of combinations of beam patterns for training in $\Omega$ is $D_B(D^{(1)}_ID^{(2)}_I)^L$ (in the order of $10^{10}$ with $L=3$ and our simulation parameters in Section \ref{sim}), and that of feasible reflection paths increases with $J$ in general. As such, the optimal beam training will incur prohibitively high (even infeasible) complexity and delay for real-time implementation, especially when the number of IRSs $J$ and/or beam patterns $D_B$/$D_I$ is practically large. A more efficient beam training approach is by sequentially updating the beam patterns at the BS and involved IRSs in $\Omega$, until the convergence is achieved. However, this also results in a high complexity in the order of $D_B+LD^{(1)}_ID^{(2)}_I$ for each reflection path $\Omega$. To avoid the above exorbitant overheads, we propose a new distributed beam training scheme, as detailed next.\vspace{-6pt}

\section{Distributed Beam Training}\label{train}
Our proposed beam training method exploits the coordination among the IRS controllers and the BS. In particular, the BS and each IRS maintain a local BRT, and all IRSs' local BRTs are fed back to the BS for facilitating the joint beamforming and beam routing design. The BRT at each IRS $j$ specifies the index of its optimal (horizontal and vertical) passive beamforming vectors in the codebook ${\cal W}_I$ (or ${\cal W}^{(1)}_I$ and ${\cal W}^{(2)}_I$) for reflecting signal from a previous node in ${\cal N}^{(P)}_j$ (BS/IRS) to its next node in ${\cal N}^{(N)}_j$ (IRS/user). Similarly, the BRT at the BS specifies the index of the optimal active beamforming vector in its codebook for transmitting to its next node in ${\cal N}_0$ (IRS). An example of the local BRT at the BS and that at IRS $j$ are shown in Fig.\,\ref{BRT}. Next, we introduce how to practically construct such BRTs at the BS and each IRS. Depending on whether the beams are active (at the BS) or passive (at each IRS), we consider the following two cases in the next two subsections, respectively.\vspace{-9pt}

\begin{figure*}
\centering
\begin{minipage}[t]{0.6\textwidth}
\centering
\includegraphics[width=4.3in]{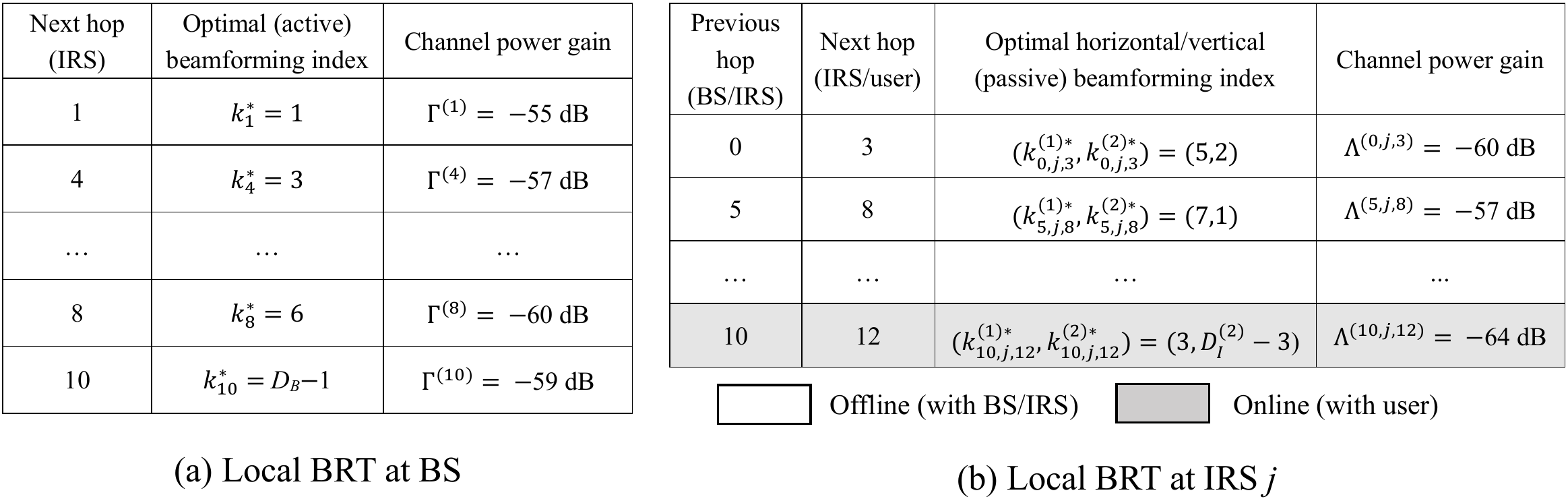}
\vspace{-15pt}
\caption{Illustrations of BRTs at the BS and IRS.}\label{BRT}
\end{minipage}
\vspace{-6pt}
\hfill
\begin{minipage}[t]{0.39\textwidth}
\centering
\includegraphics[width=2.7in]{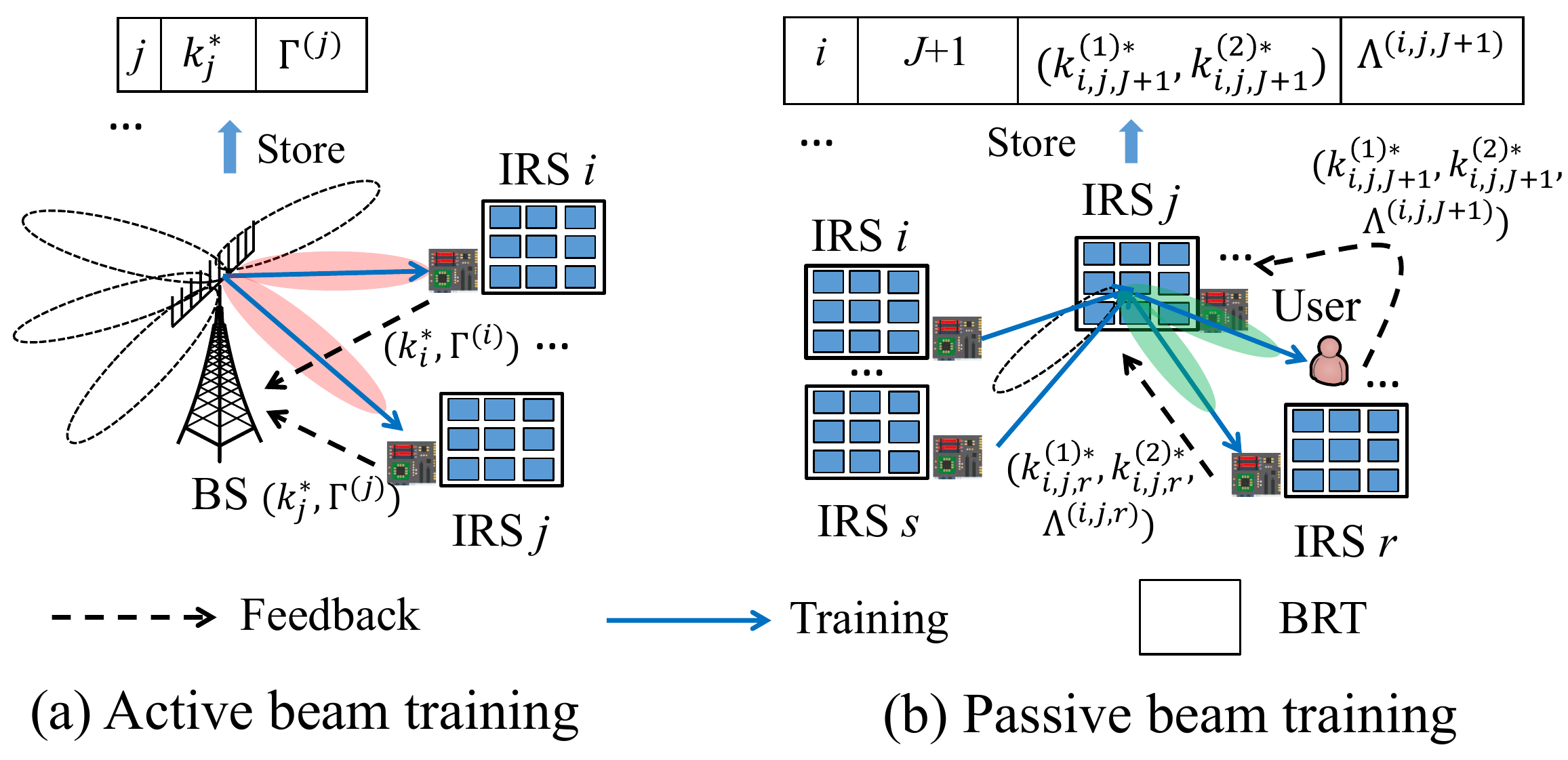}
\vspace{-6pt}
\caption{Proposed distributed beam training.}\label{train_IRS}
\end{minipage}
\vspace{-12pt}
\end{figure*}
\subsection{Active Beam Training at BS}\label{activeBT}
By conducting the active beam training offline, the BS (node 0) constructs its local BRT, as shown in Fig.\,\ref{BRT}(a). Specifically, as shown in Fig.\,\ref{train_IRS}(a), the BS consecutively sends training symbols from each direction defined in ${\cal W}_B$. In the meanwhile, each IRS controller in ${\cal N}_0$ measures its average RSS for each active beam direction. Let $\Gamma_n^j$ denote the average RSS (normalized by the BS transmit power) at IRS controller $j, j \in {\cal N}_0$ when the $n$-th beam in ${\cal W}_B$ is used by the BS. Then, the index of the optimal active beamforming vector for the BS to transmit to IRS $j$ is obtained as $k_j^*=\arg \mathop {\max}\nolimits_{1 \le k \le D_B} \Gamma_k^{(j)}$, and the maximum RSS at IRS $j$ is given by $\Gamma^{(j)}=\Gamma_{k_j^*}^{(j)}$. Next, each IRS controller $j$ in ${\cal N}_0$ reports $k_j^*$ and $\Gamma^{(j)}$ to the BS, which are stored in the local BRT at the BS as its optimal (active) beamforming index and channel power gain with IRS $j$, respectively, as shown in Fig.\,\ref{BRT}(a). \vspace{-6pt}

\subsection{Passive Beam Training at IRS}\label{ptb}
Similar to the BS, each IRS controller performs passive beam training to construct its BRT, as shown in Fig.\,\ref{BRT}(b), with the help of its neighboring nodes in ${\cal N}_j$. Depending on whether the user is involved or not (online versus offline training), we further consider the following two cases.

{\it Case 1: Offline training without the user.} As shown in Fig.\,\ref{train_IRS}(b), IRS controller $j$ informs each node $i, i \in {\cal N}^{(P)}_j$ (either an IRS controller or the BS) which is a previous hopping node of it to send training signal over consecutive time slots, while it changes the reflection direction based on ${\cal W}_I$ sequentially over time slots. Meanwhile, each IRS controller $r, r \in {\cal N}^{(N)}_j$, which is a next hopping node of it measures the average RSS with its different reflection directions, and sends back to it the best one that has the maximum RSS. Note that in the case that the previous node is the BS, i.e., $i=0$, the BS can set its active beamforming as the optimal one for IRS $j$, i.e., the $k_j^*$-th beam in ${\cal W}_B$ obtained after the active beam training, so as to enhance the RSS at each IRS. However, unlike the active beam training, the signal propagated through the direct link from node $i$ to node $r$ may interfere with the RSS measurement and thus should be eliminated by IRS controller $r$. To this end, in the passive beam training, IRS controller $j$ can first turn off IRS $j$ for node $r$ to measure the received signal only through its direct link with node $i$. Then, the direct channel component can be cancelled in the subsequent RSS measurement. By this means, IRS controller $j$ can determine the optimal passive beamforming for IRS $j$ to reflect the beam from node $i$ to node $r$ given the triple of nodes $(i,j,r), i \in {\cal N}^{(P)}_j, r \in {\cal N}^{(N)}_j$. Let $\Lambda_n^{(i,j,r)}$ denote the measured RSS from node $i$ at node $r$ when IRS $j$ uses the $n$-th beam in ${\cal W}_I$. Then, the index of the optimal passive beamforming vector for this triple of nodes $(i,j,r)$ is obtained as $k_{i,j,r}^*=\arg \mathop {\max}\nolimits_{1 \le k \le D_I} \Lambda_k^{(i,j,r)}$. 

Note that the complexity of the above 3D beam training at each IRS can be reduced by decoupling it into horizontal and vertical passive beam training. For example, IRS controller $j$ can first fix the vertical beam direction of IRS $j$ as any one in $W_I^{(2)}$ and search for its corresponding optimal horizontal beam direction in $W_I^{(1)}$. Then, it sets the horizontal beam direction as the optimal one found and searches for its corresponding optimal vertical beam direction in $W_I^{(2)}$. Although the above simplified training may not find the optimal horizontal and vertical passive beamforming pair, it significantly reduces the optimal training complexity from ${\cal O}(D_I^{(1)}D_I^{(2)})$ to ${\cal O}(D_I^{(1)}\!\!+\!\!D_I^{(2)})$ for each group of triple nodes. Denote by $k^{(1)*}_{i,j,r}$/$k^{(2)*}_{i,j,r}$ and $\Lambda^{(i,j,r)}$ the index of the obtained horizontal/vertical beamforming vector and the corresponding maximum RSS at node $r$ for the triple nodes $(i,j,r)$, respectively. Both $k^{(1)*}_{i,j,r}$/$k^{(2)*}_{i,j,r}$ and $\Lambda^{(i,j,r)}$ are collected by IRS controller $j$ and stored in its BRT as its optimal horizontal/vertical passive beamforming index and cascaded channel power gain from the previous node $i$ to the next node $r$ via IRS $j$, respectively, as illustrated in Fig.\,\ref{BRT}(b). 

{\it Case 2: Online training with the user.} If node $r$ is the user, i.e., $r=J+1$, similar beam training scheme as in Case 1 can be conducted, but in an online manner. Suppose that the user has been discovered to be a neighboring node of IRS $j$ via their online coordination. Then, IRS controller $j$ informs all nodes in ${\cal N}_j^{(P)}$ to send training signals, while the user measures $\Lambda^{(i,j,J+1)}, i \in {\cal N}_j^{(P)}$ and reports them along with $k^{(1)*}_{i,j,J+1}$ and $k^{(2)*}_{i,j,J+1}$ to IRS controller $j$, similarly as IRS controller $r$ in Case 1, as shown in Fig.\,\ref{train_IRS}(b). These parameters are then stored in the BRT at IRS controller $j$, as shown in Fig.\,\ref{BRT}(b). As such, the online complexity of the proposed beam training is in the order of \small$D_I^{(1)}+D_I^{(2)}$\normalsize per neighbor IRS of the user, as compared to \small$D_B(D^{(1)}_ID^{(2)}_I)^L$\normalsize and \small$D_B+LD^{(1)}_ID^{(2)}_I$\normalsize per reflection path in the exhaustive and sequential beam search, respectively.\vspace{-6pt}

\subsection{Multi-hop Channel Gain Estimation and Routing Design}
Each IRS controller $j, j \in \cal J$ feeds back its BRT obtained offline as well as its average channel power gains with other controllers in its neighborhood, i.e., $Q_{i,j}, i \in {\cal N}_j^{(P)}$ and $Q_{j,r}, r \in {\cal N}_j^{(N)}, r \ne J+1$, to the BS. If a user is discovered in its neighborhood, it also sends back the corresponding new entries in its BRT (see, e.g., the last row of Fig.\,\ref{BRT}(b)) obtained online to the BS. Based on such feedback as well as its own BRT, the BS can conduct a global BRT to estimate an approximate value of the maximum end-to-end channel power gain for any reflection path $\Omega$. If the BS applies the $k_{a_1}^*$-th active beam in ${\cal W}_B$ and each IRS $a_l$ applies the $k^{(1)*}_{a_{l-1},a_l,a_{l+1}}$/$k^{(2)*}_{a_{l-1},a_l,a_{l+1}}$-th horizontal/vertical passive beam in $W_I^{(1)}$/$W_I^{(2)}$, the maximum channel power gain for a given $\Omega$ is approximated as
\begin{equation}\label{approx}
	\lvert \tilde h_{0,J+1}(\Omega) \rvert^2=
\begin{cases}
	\frac{\prod\limits_{l=1}^{L}\Lambda^{(a_{l-1},a_l,a_{l+1})}}{\prod\limits_{l=1}^{L-1}Q_{a_l,a_{l+1}}},&{\text{if}}\; L \ge 2 \\[12pt]
	\Lambda^{(0,a_1,a_2)}, &{\text{if}}\; L = 1.
\end{cases}
\end{equation}
Note that each $Q_{a_l,a_{l+1}}$ in the denominator in the first case of (\ref{approx}) is used to approximate the large-scale path gain between IRSs $a_l$ and $a_{l+1}$. It can be shown that the approximate maximum channel power gain in (\ref{approx}) becomes {\it exact} if only LoS propagation exists in all constituent links of $\Omega$\cite{mei2021massive}. 

Next, to determine the beam-routing solution, the BS optimizes the reflection path $\Omega$ to maximize the end-to-end channel power gain $\lvert \tilde h_{0,J+k}(\Omega) \rvert^2$ in (\ref{approx}), i.e.,
\begin{equation}\label{op2}
\mathop {\max}\limits_{\Omega}\; \lvert \tilde h_{0,J+1}(\Omega) \rvert^2,\;\;\text{s.t.}\;\;e_{a_l,a_{l+1}} \in E_L, \forall l \in \cal L.
\end{equation}
Problem (\ref{op2}) can be viewed as a special case of the multi-beam routing problem for multiple users studied in our prior work\cite{mei2021massive}, which can be optimally solved by utilizing a graph-optimization approach. Due to the page limit, the details of solving (\ref{op2}) are omitted in this letter. Note that the proposed beam training scheme may only yield a suboptimal beam training solution, as it only accounts for partial CSI in the system involving the IRS controllers. However, as will be shown in Section \ref{sim}, it is able to yield a high-quality solution with significantly reduced complexity.\vspace{-6pt}

\section{Simulation Results}\label{sim}
In this section, we provide simulation results to evaluate the performance of our proposed distributed beam training scheme. We consider an indoor multi-IRS aided system with two typical user locations, as shown in Fig.\,\ref{topology}(a). Given the user locations, it suffices to consider $J=5$ IRSs shown in Fig.\,\ref{topology}(a). The LoS graph of this system for user location 1 is shown in Fig.\,\ref{topology}(b). Unless otherwise stated, the simulation parameters are set as follow. The number of BS antennas is $N=16$. If there is an LoS link between nodes $i$ and $j$, i.e., $e_{i,j} \in E_L$, we assume that the channel between them follows Rician fading with a Rician factor of $K=10$ dB. Otherwise, it follows Rayleigh fading with a path-loss exponent of 3.5. The carrier frequency is set to $f_c=5$ GHz. The antenna and element spacing at the BS and each IRS is set to half- and quarter-wavelength, respectively. The numbers of elements in each IRS's horizontal and vertical dimensions are equal, i.e., $M_1=M_2\triangleq M_0$. The number of beam patterns in ${\cal W}_B$ is set to 16, while that in ${\cal W}^{(1)}_I$ and ${\cal W}^{(2)}_I$ are both set to 32, i.e., $D_B=16$ and $D^{(1)}_I=D^{(2)}_I=32$. Accordingly, we use 16- and 32-point discrete Fourier transform (DFT)-based codebooks at the BS and each IRS (horizontal and vertical), respectively. All results shown in this section are averaged over 100 independent channel realizations. Due to the formidable complexity of the exhaustive beam search for the optimal beam training, we consider the sequential beam search for performance comparison.
\begin{figure}[!t]
\centering
\subfigure[3D plot.]{\includegraphics[width=0.24\textwidth]{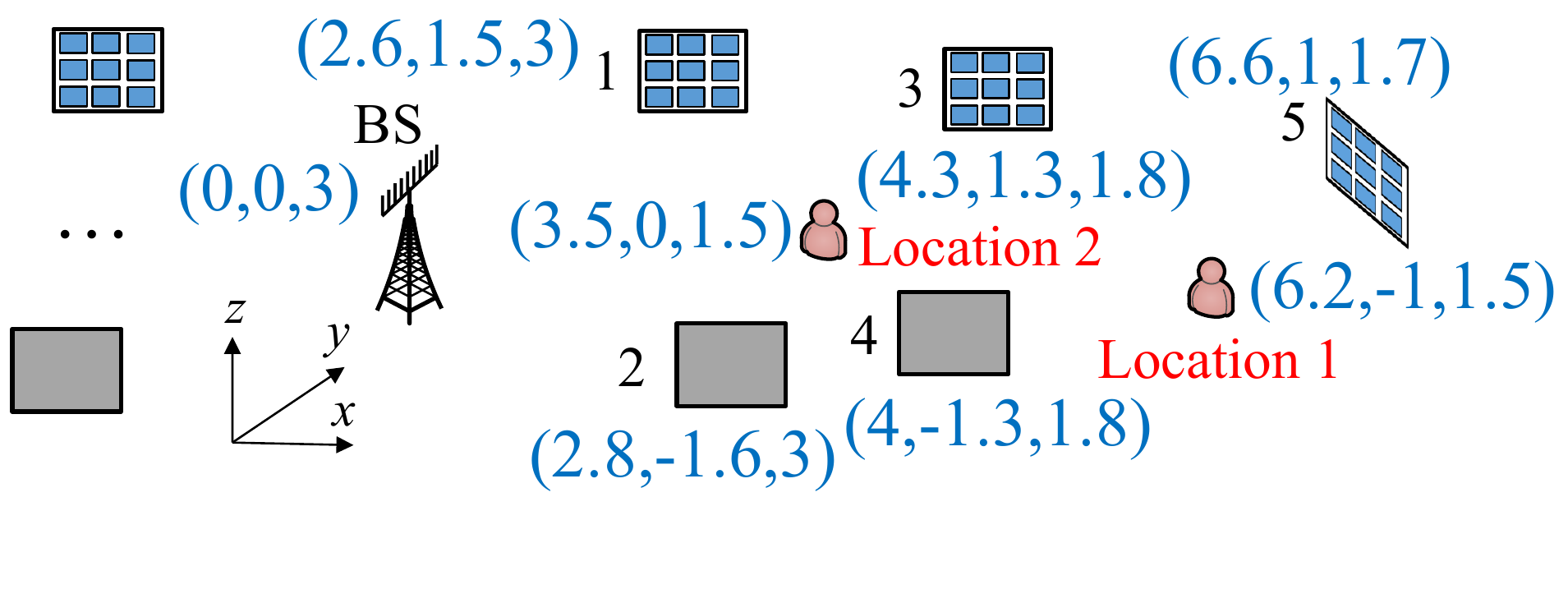}}
\subfigure[LoS graph for user location 1.]{\includegraphics[width=0.24\textwidth]{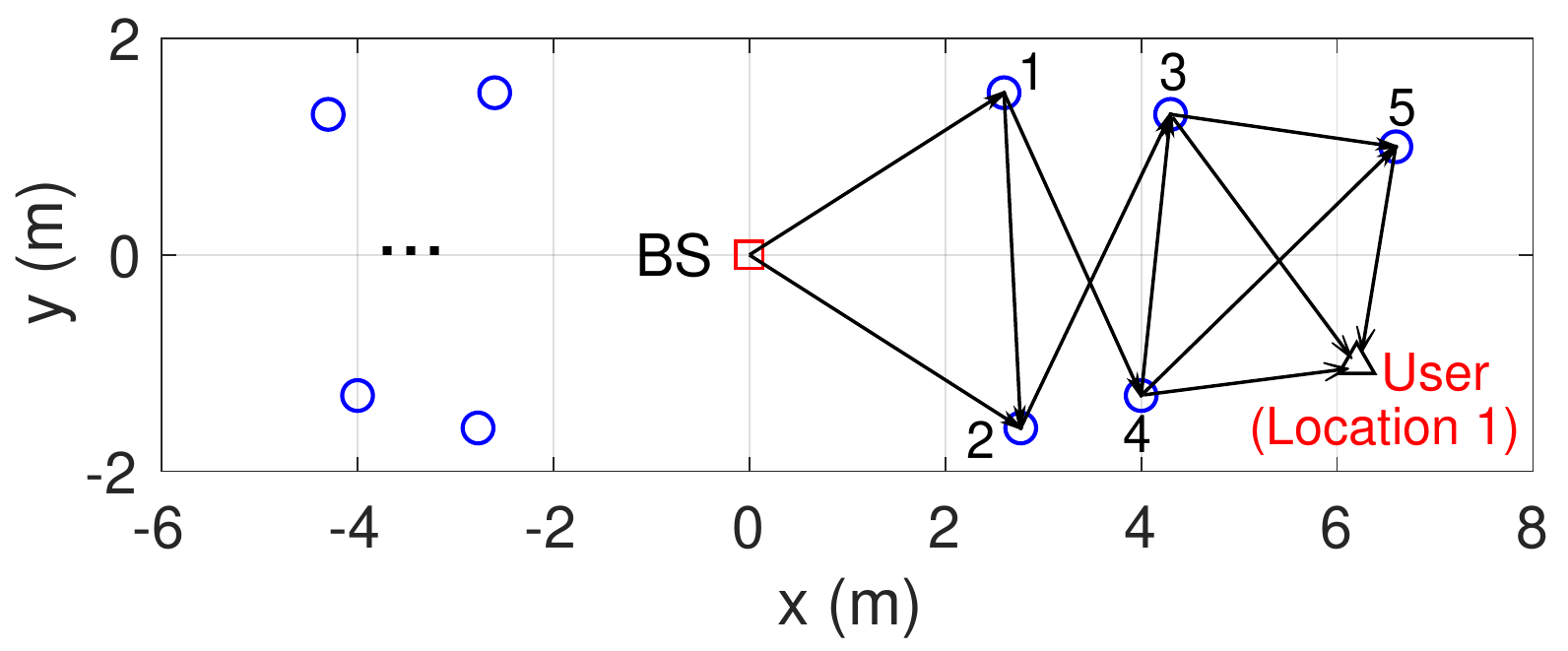}}
\vspace{-9pt}
\caption{Simulation setups.}\label{topology}
\vspace{-6pt}
\end{figure}

\begin{figure}[hbtp]
\centering
\subfigure[]{\includegraphics[width=0.24\textwidth]{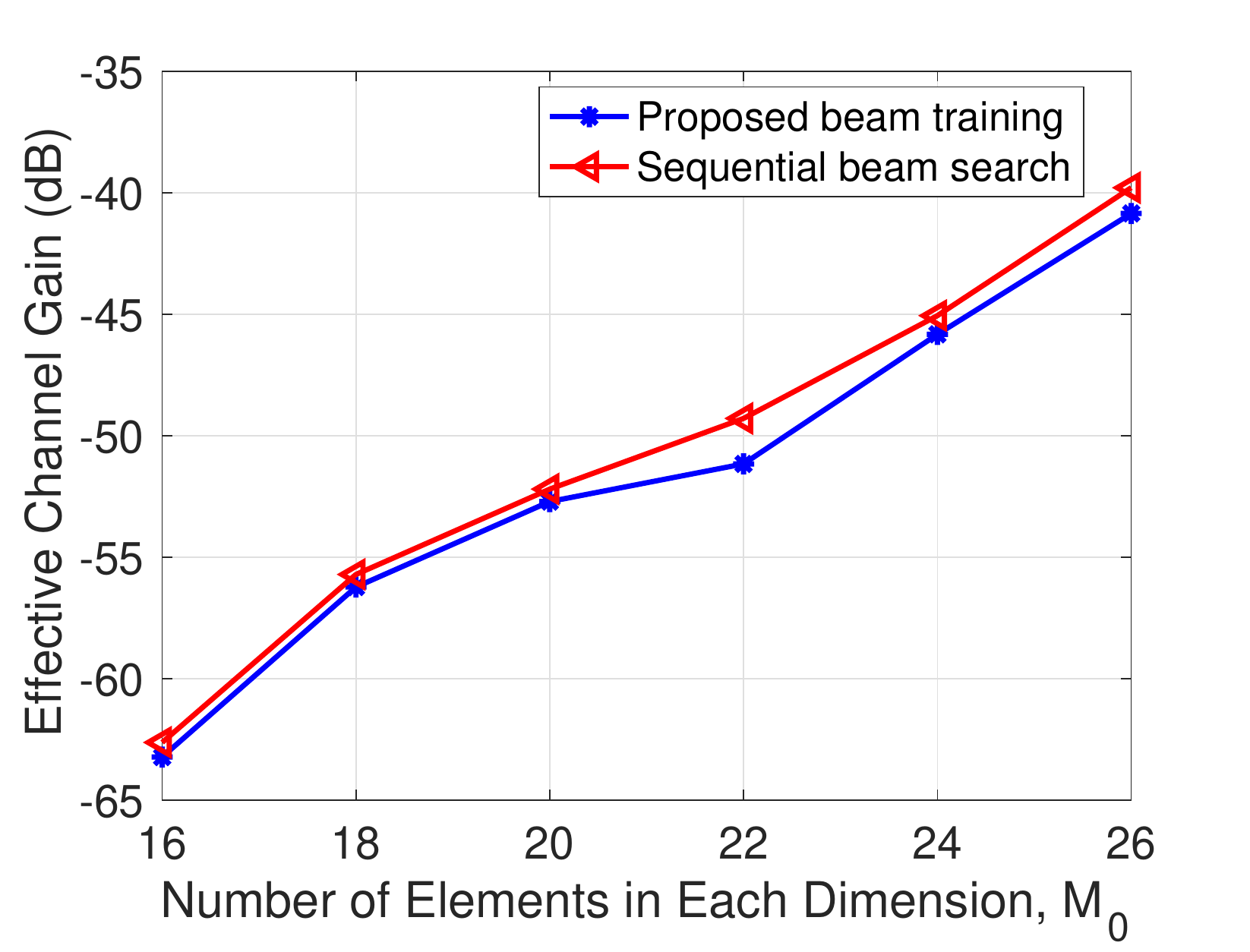}}
\subfigure[]{\includegraphics[width=0.24\textwidth]{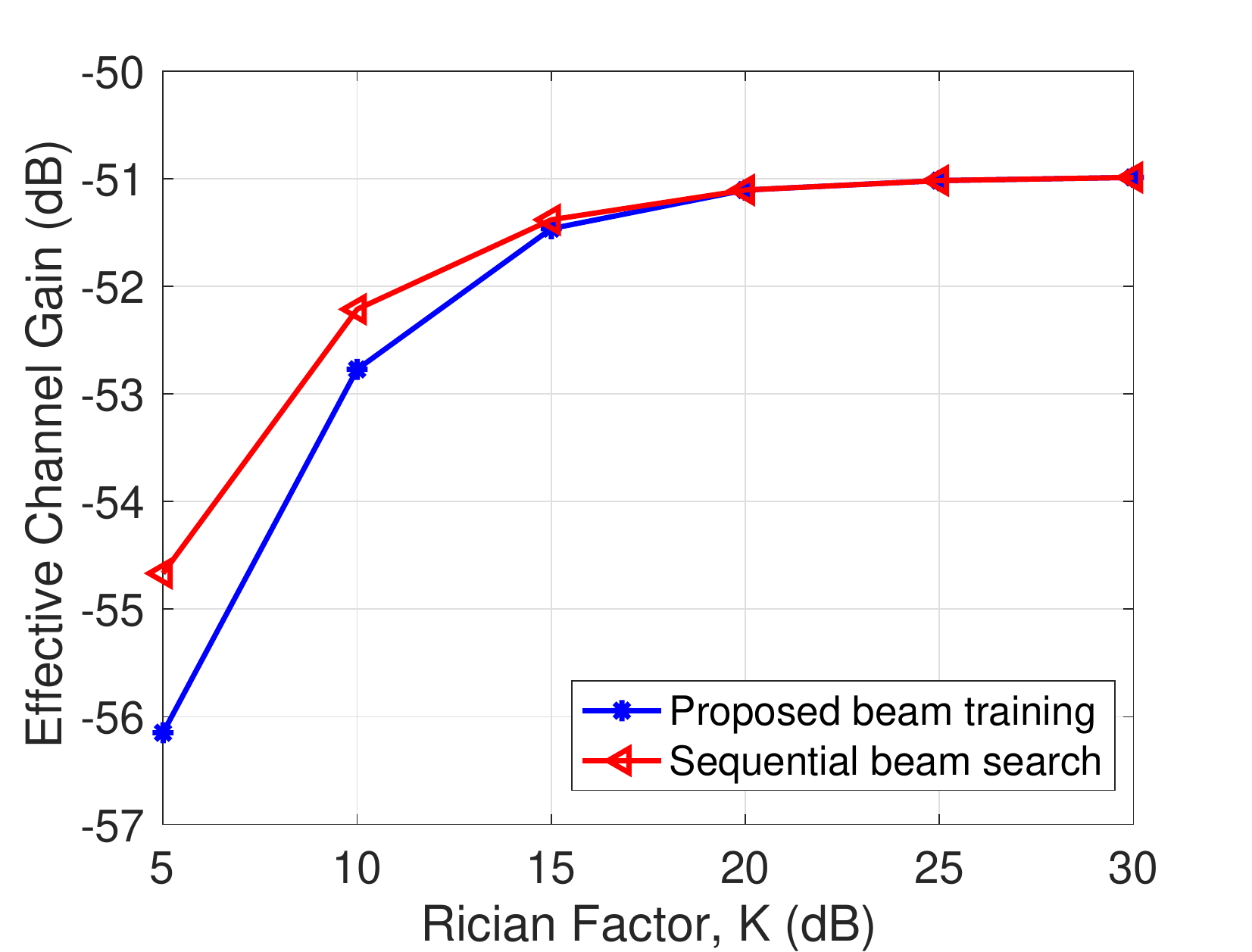}}
\subfigure[]{\includegraphics[width=0.24\textwidth]{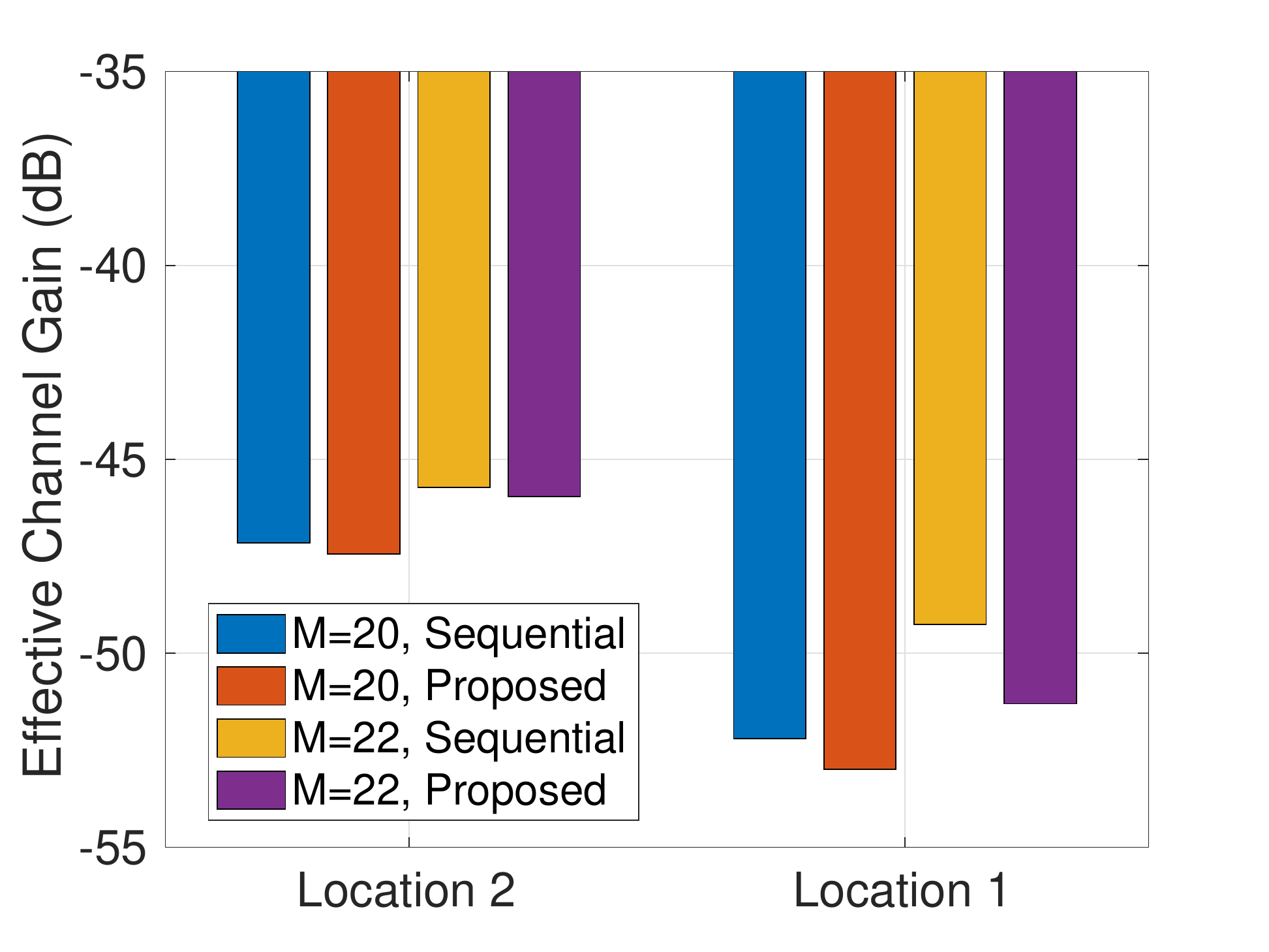}}
\subfigure[]{\includegraphics[width=0.24\textwidth]{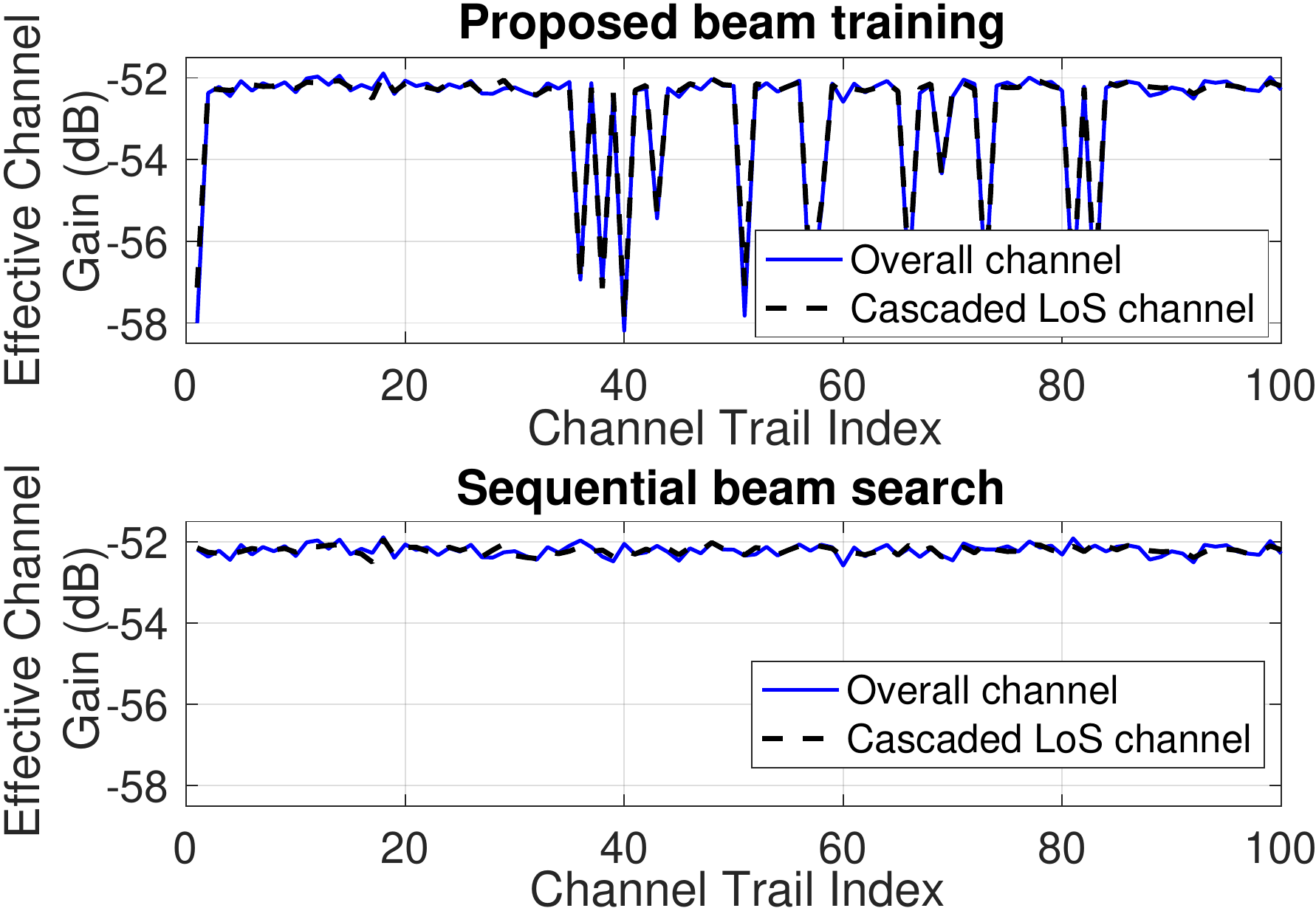}}
\vspace{-12pt}
\caption{Effective channel power gain versus (a) number of IRS elements in each dimension $M_0$, (b) Rician factor $K$, and (c) user location; (d) Cascaded LoS versus overall channel power gain realizations.}\label{simfig}
\vspace{-6pt}
\end{figure}
First, Fig.\,\ref{simfig}(a) plots the effective BS-user channel power gains at user location 1 versus $M_0$ using the proposed beam training or the sequential beam search. It is observed that their performance gap is small over the whole range of $M_0$ considered. Nonetheless, the gap becomes slightly larger as $M_0 \ge 22$. This is because the hop number of the optimized reflection path may increase with $M_0$ to achieve higher CPB gain to compensate for the path-loss\cite{mei2020cooperative}, which results in higher approximation error in our proposed distributed training. Second, Fig.\,\ref{simfig}(b) shows the effective BS-user channel power gains at user location 1 based on the two considered schemes versus the Rician factor $K$ with $M_0=20$. It is observed that they both increase with $K$, since for a given reflection path $\Omega$, the LoS component between any two consecutive nodes in $\Omega$ becomes more dominant, thus enhancing both the active beamforming and CPB gains in $\Omega$. Moreover, their performance gap is observed to decrease with $K$ and becomes negligible as $K \ge 15$ dB. This is expected since the approximation accuracy of the proposed distributed beam training improves as $K$ increases. In Fig.\,\ref{simfig}(c), we compare their performance gaps at the two user locations. It is observed that the performance gap is smaller when the user is at location 2. The reason is that location 2 is closer to the BS than location 1 and thus, the hop number of the optimized reflection path is generally smaller, which helps reduce the accumulated error of distributed training. Finally, to evaluate the strength of other randomly scattered links under the optimized beamforming and beam routing solutions by the two considered schemes, we plot their respective power gains of the cascaded LoS channel and overall channel (cascaded LoS plus scattered links) over the 100 channel trials in Fig.\,\ref{simfig}(d), with $M_0=20$. It is observed that the strength of the cascaded LoS channel is comparable to that of the overall channel in both considered schemes, which implies that the strength of the randomly scattered links is practically negligible. It is also observed that the sequential beam search yields less fluctuating channel power gains than the proposed beam training, as expected.\vspace{-3pt}

\section{Conclusions}
In this letter, we propose a new and efficient distributed beam training based design for joint active/passive beamforming and beam routing in a multi-IRS aided system, without the need of high-complexity exhaustive or sequential beam search. It is shown via simulation that the performance gap between the proposed design and the sequential beam search is practically small, thus validating the former as an appealing low-complexity solution. The proposed training method for a single user can be extended/applied to general multi-IRS aided systems with multiple users and different design objectives by constructing their corresponding BRTs.\vspace{-3pt}

\bibliography{IRStrain.bib}
\bibliographystyle{IEEEtran}
\end{document}